\documentclass[amsmath,amssymb,floatfix,showpacs,twocolumn]{revtex4}
\usepackage{amsfonts}
\usepackage{graphicx}
\usepackage{dcolumn}
\usepackage{bm}
\usepackage{hyperref} 
\usepackage{latexsym}
\usepackage{float}

\def\Pinf{P_{\infty}}

\begin{document}
\title{Percolation in Hierarchical Scale-Free Nets}   

\author{Hern\'an D. Rozenfeld}
\email{rozenfhd@clarkson.edu}
\affiliation{Department of Physics, Clarkson University,
Potsdam NY 13699-5820, USA}
\author{Daniel ben-Avraham}
\email{qd00@clarkson.edu}
\affiliation{Department of Physics, Clarkson University,
Potsdam NY 13699-5820, USA}

\begin{abstract}
We study the percolation phase transition in hierarchical scale-free nets.  Depending on the method
of construction, the nets can be fractal  or small-world (the diameter grows either algebraically or  logarithmically with the net size), assortative or disassortative (a measure of the tendency of like-degree nodes to be connected to one another), or possess various degrees of clustering.  The percolation phase transition can be analyzed exactly in all these cases, due to the self-similar structure of the hierarchical nets.  We find different types of criticality, illustrating the crucial effect of other structural properties besides the scale-free degree distribution of the nets.

\end{abstract}

\maketitle

\section{INTRODUCTION}

Many large complex nets, such as the Internet and the World Wide Web, social networks of contact,
and networks of interactions between proteins are  {\it scale-free\/}: the degree $k_i$ (number of links attached to a node, $i$) has a distribution with a heavy power-law tail, $P(k)\sim k^{-\gamma}$.  Because of their ubiquitousness in everyday life the structure and physical properties of scale-free nets have attracted much recent attention~\cite{reviews}.

The percolation problem is of particular practical interest: Is the integrity of the Internet compromised following random breakdown of a fraction of its routers? What fraction of a population ought to be vaccinated to arrest the spread of an epidemics that spreads by social contact?  Initial studies 
of percolation addressed the case of {\it stochastic\/} scale-free nets, where the links between the nodes are drawn at random, so as to satisfy the scale-free degree distribution (for example, by the algorithm due to Molloy and Reed~\cite{molloy}).  These studies showed that scale-free nets are resilient to random dilution, provided that the degree exponent $\gamma$ is smaller than 3. Explicit expressions for the critical exponents characterizing the transition as a function of $\gamma$ were also 
derived~\cite{albert,callaway,cohen,cohen2}.  

Stochastic Molloy-Reed scale-free nets are limited, though.  Having fixed the degree distribution, all other structural 
properties (such as the extent of clustering, assortativity, etc.)\ are fixed as well, in contrast with man-made and natural scale-free nets that show a great deal of variation in these other properties.
In this paper, we study percolation in 
 {\it hierarchical\/} scale-free nets~\cite{berker,kaufman,hinczewski}.  Hierarchical scale-free nets may be constructed that are small-world or not, and with various degrees  of assortativity, clustering, and other properties~\cite{rozenfeld}.  
 
Hierarchical nets have been studied before, as exotic examples where renormalization group techniques yield exact results~\cite{berker,kaufman,hinczewski}, including the percolation phase transition and the $q\to1$ limit of the Potts model~\cite{hong,andelman}.   
 We study percolation directly, by focusing on the size of the giant component (the largest component left after dilution) and the probability of contact between hubs (nodes of highest degree). Our aim is to elucidate the effect of the various structural properties of the nets on the percolation phase transition.  As we shall see below, whether the transition takes place or not, and its character, depends
 not only on the degree exponent (as in stochastic nets) but also on other factors~\cite{boguna}.

\section{HIERARCHICAL SCALE-FREE NETS}

\begin{figure}[ht]
  \vspace*{0.0cm}\includegraphics*[width=0.40\textwidth]{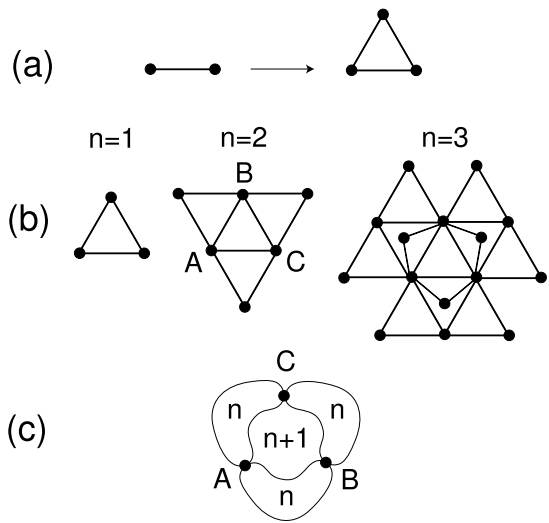}
\caption{The $(1,2)$-flower, or DGM network. 
(a)~Method of construction: each link in generation $n$ is replaced by two paths of 1 and 2 links long. (b)~Generations $n=1,2,3$.  (c)~Alternative method of construction: generation $n+1$ is obtained by joining three replicas of generation $n$ at the hubs (marked by A,B,C).
}
\label{fig1}
\end{figure}

Hierarchical scale-free nets are constructed in a recursive fashion.  We focus on the special class
of $(u,v)$-flowers~\cite{rozenfeld}, where each link in generation $n$ is replaced by two parallel paths consisting of $u$ and $v$ links, to yield generation $n+1$.  A natural
choice for the genus at generation $n=1$ is a cycle graph (a ring) consisting of $u+v\equiv w$ links and nodes (other choices are possible). The case of $u=1$, $v=2$ (Fig.~\ref{fig1}) has been studied previously by Dorogotsev, Goltsev and Mendes (DGM)~\cite{dorogovtsev}.  In the following we assume that $u\leq v$, without loss of generality.

 All $(u,v)$-flowers are self-similar, as evident from an equivalent method of construction: to produce generation $n+1$, make $w$ copies of the net in generation $n$ and join them at the hubs (the nodes of highest degree), as illustrated in Fig.~\ref{fig1}c.

It is easy to see, from the second method of construction, that the number of links (the size) of a $(u,v)$-flower of generation $n$ is
\begin{equation}
\label{Mn}
M_n=(u+v)^n= w^n.
\end{equation}
At the same time, the number of nodes (the order) obeys the recursion relation
\[
N_n=wN_{n-1}-w\,,
\]
which, together with the boundary condition $N_1=w$, yields
\begin{equation}
\label{Nn}
N_n=\Big(\frac{w-2}{w-1}\Big)w^n+\Big(\frac{w}{w-1}\Big)\,.
\end{equation}

Similar considerations let us reproduce the full degree distribution.  By construction, $(u,v)$-flowers have only nodes of degree $k=2^m$, $m=1,2,\dots,n$.  Let $N_n(m)$ be the number of nodes of degree $2^m$
in the $(u,v)$-flower of generation $n$, then
\[
N_n(m)=N_{n-1}(m-1)+(w-2)w^{n-1}\delta_{m,1}\,,
\]
leading to
\begin{equation}
\label{Nk}
N_{n}(m) = \left\{ \begin{array}{ll}
(w-2)w^{n-m}, \quad  & m<n,\\
w \quad & m=n.
\end{array} \right.
\end{equation}
As in the DGM case, this corresponds to a scale-free degree distribution, $P(k)\sim k^{-\gamma}$,
of degree exponent
\begin{equation}
\label{gamma}
\gamma=1+\frac{\ln(u+v)}{\ln2}\,.
\end{equation}

The self-similarity of $(u,v)$-nets, coupled with the fact that different replicas meet at a {\it single\/} 
node, makes them amenable to exact analysis by renormalization group techniques.  

\subsection*{Network diameter and dimensionality}
There is a vast difference between $(u,v)$-flowers with $u=1$ and $u>1$.
If $u=1$ the diameter $L_n$ of the $n$-th generation flower (the longest shortest path between any two nodes) scales linearly with $n$.  For example, $L_n=n$ for the $(1,2)$-flower~\cite{dorogovtsev} and
$L_n=2n$ for the $(1,3)$-flower.  It is easy to see that the diameter of the $(1 ,v)$-flower, for $v$ odd, is
$L_n=(v-1)n+(3-v)/2$, and, while deriving a similar result for $v$ even is far from trivial, one can show that $L_n\sim(v-1)n$.

For $u>1$, however, the diameter grows as a power of $n$.  For example, for the $(2,2)$-flower we find
$L_n=2^n$, and, more generally, if $u+v$ is even (and $u>1$),
\[
L_n = \biggl(\frac{u+v}{2} + \frac{v-u}{u-1}\biggr) u^{n-1} - \frac{v-u}{u-1}\,.
\]
For $u+v$ odd one may establish bounds 
showing that $L_n\sim u^n$.
To summarize,
\begin{equation}
\label{Ln}
L_n \sim \left\{ \begin{array}{ll}
(v-1)n &  u=1,\\
u^{n} & u > 1.
\end{array} \right.
\end{equation}

Since $N_n\sim (u+v)^n$, we can recast these relations as
\begin{equation}
\label{L}
L \sim \left\{ \begin{array}{ll}
\ln N &  u=1,\\
N^{\ln u/\ln(u+v)} & u > 1.
\end{array} \right.
\end{equation}
Thus, for $u=1$ the flowers are {\it small world\/}, similar to stochastic scale-free nets with $\gamma>3$.
For $u>1$ the nets are in fact {\it fractal\/}, with fractal dimension
\begin{equation}
d=\frac{\ln(u+v)}{\ln u}\,,\qquad u>1\,,
\end{equation} 
since the mass increases by $u+v$ (from one generation to the next) while the diameter increases
by $u$.  $(1,v)$-flowers are {\it infinite\/}-dimensional.  In~\cite{rozenfeld} we showed how these nets may
be characterized by a different measure of dimension that takes into account their small-world scaling.

\begin{figure}[ht]
  \vspace*{0.0cm}\includegraphics*[width=0.40\textwidth]{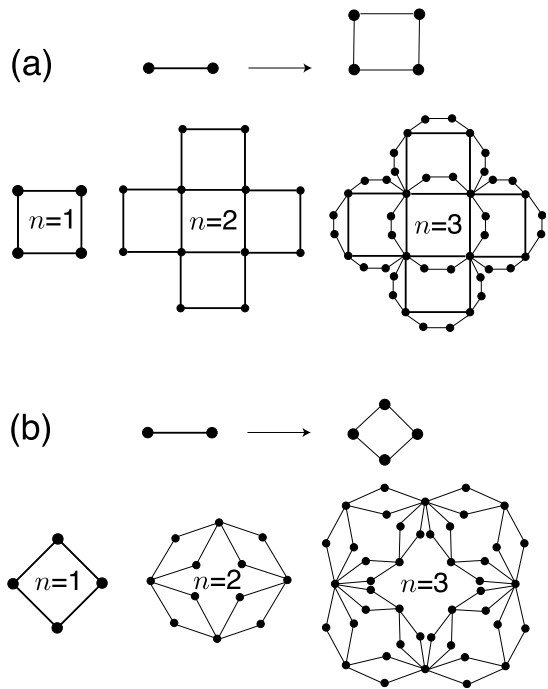}
\caption{$(u,v)$-flowers with $u+v=4$ ($\gamma=3$).
(a)~Small world: $u=1$ and $v=3$. (b)~Fractal: $u=2$ and $v=2$.  The graphs may also
be iterated by joining four replicas of generation $n$ at the appropriate hubs.
}
\label{fig2}
\end{figure}

The difference between flowers with $u=1$ and $u>1$ is perhaps best exemplified by the (1,3)- vs.\ the (2,2)-flower (Fig.~\ref{fig2}).  The nets have identical degree distributions, node for node, with degree
exponent $\gamma=3$ --- similar to the famed Barab\'asi-Albert model~\cite{barabasi} --- but the 
$(1,3)$-flower is
small world (or infinite-dimensional), while the $(2,2)$-flower is a fractal of dimension $d=2$.

\subsection*{Other structural properties}

Upon varying $u$ and $v$ the hierarchical flowers acquire different structural properties.
Consider, for example, their {\it assortativity\/} --- 
the extent to which nodes of similar degree connect with one 
another~\cite{assortativity}.  In the $(1,3)$-flower, nodes of degree $2^m$ and $2^{m+1}$ are only {\it one\/} link apart, 
and the assortativity index is 0; while in the $(2,2)$-flower the same nodes are $2^{m-1}$ links apart, and
its assortativity index tends to $-1/2$ (as $N\to\infty$), indicating a high degree of disassortativity, and more in line with naturally occurring scale-free nets~\cite{Song2,strogatz}.  
More generally, the degree of assortativity, $r$, is $r\to(v-3)/2v$,  for $u=1$,
and $r\to-2/(u+v)$,  for $u>1$, (as $N\to\infty$)~\cite{rozenfeld,thesis}.

Another property of interest is {\it clustering\/}, a measure of the likelihood for neighbors of a node 
to be neighbors of one another~\cite{watts}.  $(u,v)$-flowers with $u>1$ have zero clustering: the neighbors of a node are {\it never\/ } neighbors of one another.  The DGM net ($u=1$, $v=2$) has clustering coefficient $c\approx0.7998$, and $c$ gets smaller with increasing $v$  (or degree exponent $\gamma$), quite in line with
the clustering coefficient of everyday life scale-free nets.

\subsection*{Decorated flowers}

\begin{figure}[ht]
  \vspace*{0.0cm}\includegraphics*[width=0.40\textwidth]{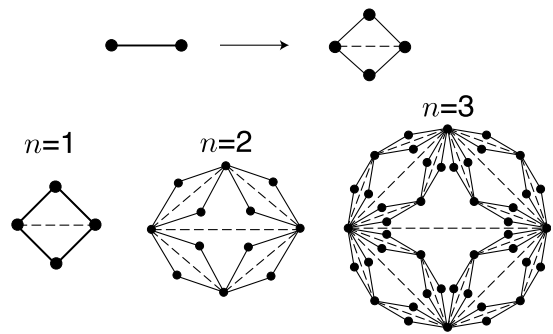}
\caption{Decorated $(2,2)$-flower.  Construction method (top): Each link is replaced by two parallel paths of $u$ and $v$ links long, and an additional link (broken line) that does not get iterated.  
(bottom):~The decorated $(2,2)$-flower for generations $n=1,2,3$.
}
\label{fig3}
\end{figure}

So far we have seen hierarchical nets that are either fractal and disassortative ($u>1$), or small world
and assortative ($u=1$ and $v>3$).  It is also possible to obtain hierarchical nets that are small world
and disassortative at the same time.  One way to do this is by constructing a fractal $(u,v)$-flower ($u>1$) and adding a link between opposite hubs at the end of each iteration step: the additional link does not get iterated~\cite{berker,kaufman,hinczewski}.  Fig.~\ref{fig3} illustrates this procedure for the case of the $(2,2)$-flower.  

The added link has a negligible effect on the degree distribution: the degree exponent still approaches
$\gamma=1+\ln(u+v)/\ln2$, as $n\to\infty$.  On the other hand, it has a dramatic effect on the diameter of the net, which now grows linearly in $n$ (logarithmically in $N$), making it a small world network.
The nets remain strongly disassortative: for example, for the $(2,2)$-flower the assortativity changes from $-1/2$ (without) to $-9/26$ with the addition of the non-iterated links.
Finally, the added links have a dramatic effect on the clustering of the nets,
which grows from zero to about 0.82008, for the $(2,2)$-flower~\cite{hinczewski}.  

The main point is that by manipulating the method of construction one can generate scale-free nets with differing structural properties.  By changing one property at a time one can then hope to understand their effect on various physical phenomena, such as the percolation phase transition.
One can also construct nets that mimic everyday life networks as closely as possible.  The decorated
$(2,2)$-flower, with its degree exponent $\gamma=3$, disassortativity, and high degree of clustering,
is a reasonable candidate for the latter.

\section{THE PERCOLATION PHASE TRANSITION} 

We now turn to the study of percolation in hierarchical scale-free nets.  The recursive nature of the $(u,v)$-flowers, coupled with their finite ramification, make it possible to obtain an exact solution by a real-space renormalization group analysis, including the finite-size behavior around the transition point.

Our plan is as follows.  We first study percolation in {\it fractal\/} hierarchical nets.  Having finite dimensionality they resemble regular and fractal lattices, and the percolation phase transition is similar
to what is found there as well.  We then study percolation in the $(1,v)$-flowers, which are small world,
as most everyday life complex networks.  Unlike everyday life nets, the $(1,v)$-flowers have no percolation phase transition, even for $v>3$, or $\gamma>3$.  Clearly, the $(1,v)$-flowers fail to
mimic everyday life networks in some crucial aspect --- perhaps their high assortativity.  We therefore conclude with an analysis of the decorated $(2,2)$-flower.  The transition there most closely resembles
that of everyday life nets, but some differences remain.  We speculate on the missing ingredient 
that gives rise to that difference in Section~\ref{discussion}.

\subsection{Fractal Scale-Free Nets}

Consider the $(2,2)$-flower, as a prototypical example of {\it fractal\/} hierarchical scale-free nets.   
In this net the distance between opposite hubs  (or the diameter) scales as $L_n\sim2^n$, and the mass scales as $N_n\sim4^n\sim L_n^2$,  corresponding to a fractal dimension of $d=2$.  Suppose that each link is present with probability $p$. Denote the probability for two opposite hubs in generation $n$  to be connected
by  $P(p)$,  then $P'$, the analogous quantity in generation $n+1$, is
\begin{equation}
\label{P22flower}
P'=2P^2-P^4.
\end{equation}
Indeed, on iterating the flower to generation $n+1$ the probability of contact between opposite hubs depends on the existence of either of two parallel paths, each consisting of two stringed copies
of generation $n$.  The probability that one of the paths is connected is $P^2$ and~(\ref{P22flower}) follows: a naive addition $P^2+P^2$ over-counts the event that all 4 generation-$n$ units are connected (probability $P^4$).

\begin{figure}[ht]
  \vspace*{0.0cm}\includegraphics*[width=0.40\textwidth]{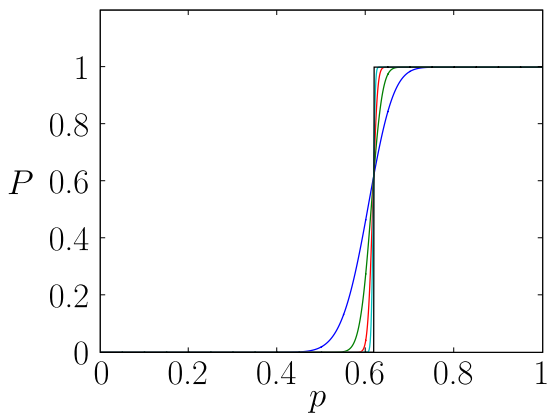}
\caption{(Color online) Contact probability between opposite hubs in the $(2,2)$-flower.  Shown are the curves
for generations $n=6,8,10,12,$ and $\infty$.
}
\label{fig4}
\end{figure}

Starting with $P=1$, for generation $n=0$, one can then compute the probability of contact $P$ for any other generation (Fig.~\ref{fig4}).  Eq.~(\ref{P22flower}) has an unstable fixed point at $p_c=(\sqrt{5}-1)/2\approx0.618$~\cite{andelman},
where $P'(p_c)=P(p_c)=p_c$, and two stable fixed points at $P=0$ and $P=1$.  If $p>p_c$, the contact probability flows to $P=1$ (as $n\to\infty$) and the system is in the percolating phase.  For $p<p_c$, it flows to $P=0$ and there is no percolation.

Near the percolation phase transition the contact probability obeys the {\it finite-size\/} scaling relation
\begin{equation}
\label{Pscaling_frac}
P=\Phi\Big(\Delta p\,L^{1/\nu}\Big),
\end{equation}
where $\Delta p=p-p_c$, 
and $\nu$ is the critical exponent governing the scaling of the correlation length, 
\begin{equation}
\xi\sim|p-p_c|^{-\nu}.
\end{equation}
We can obtain $\nu$ by evaluating the derivative of Eq.~(\ref{P22flower}) at $p=p_c$:
\begin{equation}
\label{P'P}
\frac{\partial P'}{\partial p}\Big|_{p_c}=\Lambda\frac{\partial P}{\partial p}\Big|_{p_c},
\end{equation}
where 
\begin{equation}
\label{Lambda}
\Lambda=\frac{\partial P'}{\partial P}\Big|_{p_c}.
\end{equation}
Using (\ref{Pscaling_frac}) it then follows that
\begin{equation}
\label{1/nu}
\Lambda=(L_{n+1}/L_n)^{1/\nu}.
\end{equation}
In our case $\Lambda=4(p_c-p_c^3)=6-2\sqrt{5}$ and $L_{n+1}/L_n=2$, yielding $\nu=1.63528\dots$

Next, we address the probability that a site belongs to the infinite incipient cluster (or the giant component), $\Pinf(p)$, in generation $n$.  It obeys the scaling relation
\begin{equation}
\label{Pinf_scaling}
\Pinf=N^{-\theta}\Psi(\Delta p\,N^{\theta/\beta}).
\end{equation}
The finite-size scaling exponent $\theta$ characterizes the size of the giant component at the transition point, $p=p_c$: $N_g\sim N^{1-\theta}$.  The scaling function has a non-analytic part 
 $\Psi(x)\sim x^{\beta}$, for small $x$, so that near the transition point 
\begin{equation}
\Pinf\sim(p-p_c)^{\beta}.
\end{equation}

\begin{figure}[ht]
  \vspace*{0.0cm}\includegraphics*[width=0.40\textwidth]{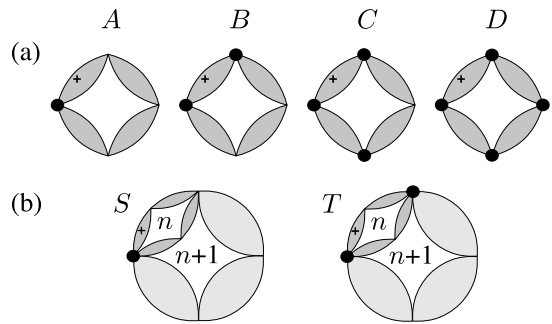}
\caption{Contact probabilities used in the derivation of Eq.~(\ref{ABCD22flower}).  See text.}
\label{fig5}
\end{figure}

Let $A,B,C$, and $D$ denote the probabilities that a site is connected to exactly one, two, three, or four of the hubs, respectively (Fig.~\ref{fig5}a), then $\Pinf=A+B+C+D$.
The analogous quantities in generation $n+1$, are
\begin{equation}
\label{ABCD22flower}
\begin{split}
& A'=SQ,\\
& B'=SPQ+TQ^2,\\
& C'=SP^2Q+T(2PQ^2),\\
& D'=SP^3+T(3P^2Q+P^3).
\end{split}
\end{equation}
Here $Q=1-P$ is the probability that opposite hubs (in generation $n$) are disconnected.  
$S$ ($T$) 
denote the event that only one (two) of the hubs that the site reaches in generation $n$ are also  hubs of generation $n+1$ (Fig.~\ref{fig5}b).  These are straightforwardly related to the $A,B,C,D$:
\begin{equation}
\label{ST22flower}
\begin{split}
& S=\frac{1}{2}A+B+\frac{1}{2}C\,,\\
& T=\frac{1}{2}C+D\,.
\end{split}
\end{equation}
As a useful check, one may verify that $\Pinf'=A'+B'+C'+D'=S+T$.
From (\ref{ABCD22flower}) and (\ref{ST22flower}) we obtain a recursion relation for $S$ and $T$:
\begin{equation}
\label{eq18}
\left(
\begin{array}{c}
S'\\
T'
\end{array}\right)=\left(
\begin{array}{cc}
\frac{1}{2}Q(1+P)^2 & Q^2(1+P)\\
\frac{1}{2}P^2(1+P) &P+P^2-P^3
\end{array}\right)
\left(
\begin{array}{c}
S\\
T
\end{array}\right)
\end{equation}
The scaling of the giant component is dominated by  $\lambda$, the largest eigenvalue of the above matrix, evaluated at $p=p_c$,
\begin{equation}
\label{lambda}
(N'/N)^{-\theta}=\lambda.
\end{equation}
In our case $\lambda=\big(7-2\sqrt{5}+\sqrt{73-32\sqrt{5}}\,\big)/4$ and $N'/N=4$, yielding $\theta=0.0503564\dots$

To obtain $\beta$, we derive Eq.~(\ref{Pinf_scaling}) with respect to $p$,
\[
\frac{\partial\Pinf}{\partial p}\Big|_{p_c}=N^{\theta\frac{1-\beta}{\beta}}\frac{\partial}{\partial p}\Psi(0)
\sim n\lambda^{n-1}\frac{\partial\lambda}{\partial p}\Big|_{p_c}\frac{\partial P}{\partial p}\Big|_{p_c},
\]
where we used the fact that $\Pinf(0)\sim\lambda^n$.  Doing the same for $P'$ and dividing the two
relations, while using~(\ref{P'P}), we get
\begin{equation}
\label{mess}
(N'/N)^{\theta\frac{1-\beta}{\beta}}\to\lambda\Lambda,\qquad\text{as $n\to\infty$}.
\end{equation}
Substituting for the values of $\lambda$, $\Lambda$, $N'/N$, and $\theta$, we find for the $(2,2)$-flower
$\beta=0.164694\dots$

\begin{figure}[ht]
  \vspace*{0.0cm}\includegraphics*[width=0.40\textwidth]{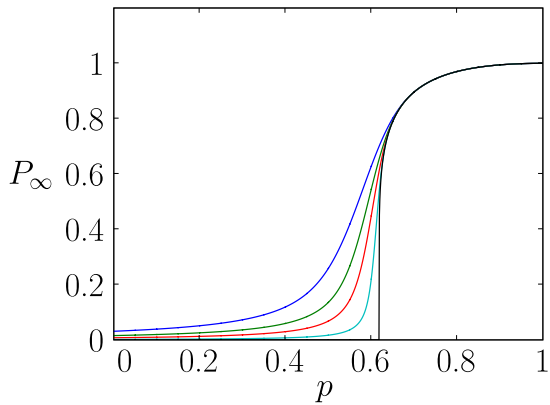}
\caption{(Color online) $\Pinf$ for the $(2,2)$-flower.  Shown are curves for generations $n=6,7,8,10,$ and $\infty$,
obtained from $\Pinf=S+T$ and iterating Eqs.~(\ref{eq18}) and (\ref{P22flower}).
}
\label{fig6}
\end{figure}

In summary, percolation in fractal scale-free nets is very similar to percolation in fractal and regular lattices.  As far as we can tell, the broad scale-free degree distribution does not give rise to
any kind of anomalous behavior, different from percolation in regular spaces.  Note, for example, that the contact probability, $P$ (Fig.~\ref{fig4}), and $\Pinf$ (Fig.~\ref{fig6}) follow the same pattern as for percolation in regular spaces.   Interestingly, the exponents
$\nu=1.63528$ and $\beta=0.164694$ that we find for the $(2,2)$-flower (whose fractal dimension is $d=2$) are not very different from the corresponding exponents in regular 2-dimensional space: 
$\nu=1.5076$ and $\beta=5/36=0.13888$.

The scaling relations between critical exponents, familiar from percolation in regular and fractal spaces,
are obeyed as well.  Indeed, we may rewrite Eq.~(\ref{1/nu}) as 
\[
\Lambda=(N'/N)^{1/\nu d}\,
\]
since $d$ is the fractal dimension of the hierarchical flower and $N\sim L^d$.  Using this, in 
conjunction with Eqs.~(\ref{lambda}) and (\ref{mess}), we derive the scaling relation
\begin{equation}
\label{thetanud}
\theta=\frac{\beta}{\nu d}\,.
\end{equation}
The giant component, at criticality, scales as 
\[
N_g\sim L^{d_g}\sim N^{d_g/d}\,,
\]
where $d_g$ is its fractal dimension.  Comparing this to $N_g\sim N^{1-\theta}$, on the one hand,
and to Eq.~(\ref{thetanud}), on the other hand, we get
\begin{equation}
\label{dg}
d_g=d-\frac{\beta}{\nu}\,,
\end{equation}
which is a well-known scaling relation for percolation in regular space~\cite{perco,remark}.

\begin{figure}[ht]
  \vspace*{0.0cm}\includegraphics*[width=0.45\textwidth]{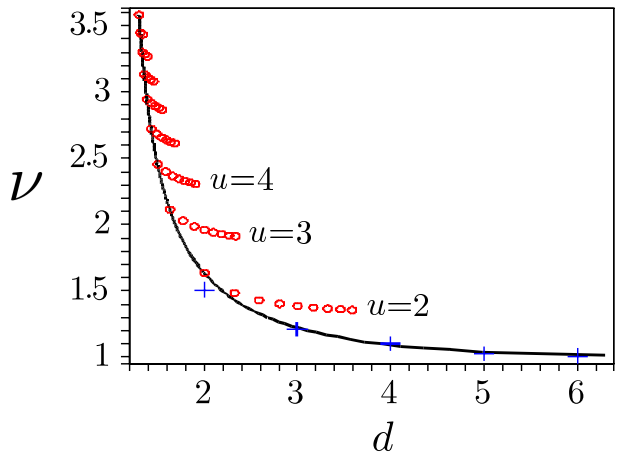}
\caption{(Color online) Correlation length exponent $\nu$ for fractal $(u,v)$-flowers ($u>1$), plotted
against their dimensionality $d=\ln(u+v)/\ln u$ ($\circ$). Shown are results for increasing
$u$ (bottom to top) and increasing $v$ (left to right).  The solid curve corresponds to $v=u$ --- a
case that is close, numerically, to regular $d$-dimensional space ($+$).
}
\label{fig7}
\end{figure}

The analysis carried out
above for the $(2,2)$-flower may be extended for other values of $u>1$ and $v$.  The recursion
relation between opposite hubs in the general case is
\begin{equation}
P'=P^u+P^v-P^{u+v}\,.
\end{equation}
As for the $(2,2)$-flower, this has two stable fixed points at $P=0$ and 1, and an unstable fixed
point  whose location at $P=p_c$ may be computed numerically.  One can then evaluate the 
correlation length exponent $\nu$, using Eqs.~(\ref{Lambda}), (\ref{1/nu}) and keeping in mind that
$L_{n+1}/L_n\to u$ 
in the thermodynamic limit.   Results from such calculations are shown in Fig.~\ref{fig7}.  There is 
general agreement with percolation in regular $d$-dimensional lattices, especially for the particular case of $v=u$ (note the analytical continuation to non-integer values of $u$ and $v$).

\subsection{Small-World, Assortative Nets}

We next turn to small world nets, and consider the $(1,v)$-flower as examples of  this type of networks.
The mass  of the $(1,v)$-flower grows like $N_n\sim(1+v)^n$, while the diameter increases 
only logarithmically, $L_n\sim\ln N_n$, making it a small world net of infinite dimensionality.
As we shall shortly see, there is no percolation transition, contradicting the finding for percolation in random scale-free nets~\cite{albert,callaway,cohen,cohen2}.   This may be perhaps attributed to the fact
that $(1,v)$-flowers are quite strongly assortative (highly connected nodes tend to be connected to
one another), making them particularly resilient to random dilution.   

The recursion relation for the probability of contact between hubs in successive generations is now
\begin{equation}
\label{eq24}
P'=P+P^v-P^{v+1}\,,
\end{equation}
which has an unstable fixed point at $p=0$ and a stable fixed point at $p=1$.  In other words, regardless
of the dilution level, $p$, contact between hubs is guaranteed, in the thermodynamic limit of $N\to\infty$.

\begin{figure}[ht]
  \vspace*{0.0cm}\includegraphics*[width=0.40\textwidth]{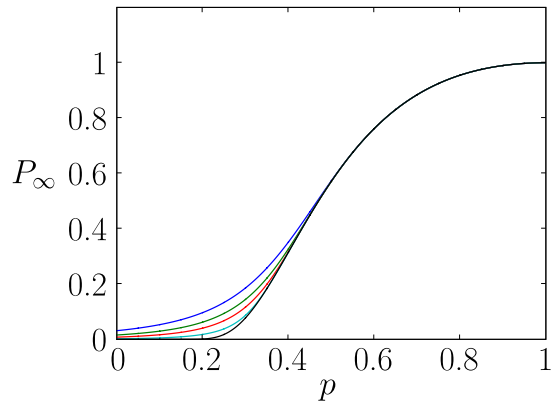}
\caption{(Color online) $\Pinf$ for the $(1,3)$-flower.  Shown are curves for generations $n=6,7,8,10$, and $\infty$,
obtained from $P=\sum_iA_i$ and iterating Eqs.~(\ref{eq24}) -- (\ref{eq26}).
Note that $\Pinf$ is indistinguishable from zero, in the scale of the plot, for $p$ below about 0.2.}
\label{fig8}
\end{figure}

Let $A_i$ ($i=1,2,\dots, v+1$) denote the probability that a node is connected to exactly $i$ adjacent hubs, in generation $n$. The recursion
relations for the analogous quantities in generation $n+1$ are
\begin{equation}
\begin{split}
&A_1'=SQ\,,\\
&A_2'=SPQ+TQ^2\,,\\
&\>\>\>\>\>\>\>\>\vdots\\
&A_i'=SP^{i-1}Q+T(i-1)P^{i-2}Q^2\,,\\
&\>\>\>\>\>\>\>\>\vdots\\
&A_{v+1}'=SP^v+T(vP^{v-1}Q+P^v)\,,
\end{split}
\end{equation}
where
\begin{equation}
\label{eq26}
\begin{split}
&S=\frac{2}{v+1}\Big(A_1+A_2+\cdots+A_v\Big)\,,\\
&T=\frac{1}{v+1}\Big(A_2+2A_3+\cdots+(v-1)A_v+(v+1)A_{v+1}\Big)\,,
\end{split}
\end{equation}
have the same meaning as for the $(2,2)$-flower, in the previous section.
Analysis of these equations reveals that there is a {\it finite\/} probability $\Pinf$ for a site to belong to the
giant component, at {\it any\/} dilution level $p$.  For $p$ small,
\begin{equation}
\label{essential}
\Pinf\sim\Big(\frac{1}{v+1}\Big)^{1/p^{v-1}}\,,
\end{equation}
and there is an essential singularity at $p=0$.  Practically, though, it is impossible to tell whether
$\Pinf=0$ or not, for $p$ sufficiently small, and one could not rule out a percolation phase 
transition at some $p_c>0$ based on a numerical study or on simulations alone (Fig.~\ref{fig8}).

\subsection{Small-World, Disassortative Nets}
Having failed to find a percolation transition in the assortative $(1,v)$-flowers, 
we now turn to the $(2,2)$-flower with a non-iterated link (Fig.~\ref{fig3}).
The recursion relation for the probability of contact between hubs in successive generations is 
\begin{equation}
\label{P'p}
P'=1-(1-p)(1-P^2)^2\,.
\end{equation}
Indeed, note that contact can be made through either of the two paths consisting of two stringed copies of generation $n$ (with probability $P^2$, in either case)  or through the non-iterated link 
(with probability $p$). The probability that none of these three parallel paths make
contact is therefore $(1-p)(1-P^2)^2$, and $P'$ follows. 

\begin{figure}[ht]
  \vspace*{0.0cm}\includegraphics*[width=0.40\textwidth]{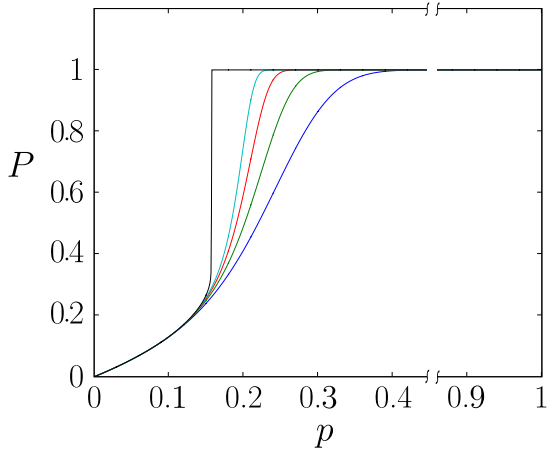}
\caption{(Color online) Contact probability between opposite hubs for the decorated $(2,2)$-flower.  Shown are
curves for generations $n=6,8,10,12,$ and $\infty$, obtained from iteration of Eq.~(\ref{P'p}).
}
\label{fig9}
\end{figure}

In the thermodynamic limit, $P'\to P$.  It is easier to obtain $P(p)$ implicitly, inverting~(\ref{P'p}):
\[
p=1-\frac{1}{(1+P)(1-P^2)}\,.
\]
One can thus see that $P(p)$ is double-valued, for $p\leq5/32$.  A stability analysis reveals that
only the lower branch is stable.  For $p>5/32$, the only available solution to (\ref{P'p}) is $P'=P=1$.
This solution is stable as well.  Thus, $P(p)$ has a discontinuity at $p_c=5/32$, where it jumps 
from $P(p_c^-)=1/3$ to $P(p_c^+)=1$, see Fig.~\ref{fig9}.   

\begin{figure}[ht]
  \vspace*{0.3cm}\includegraphics*[width=0.40\textwidth]{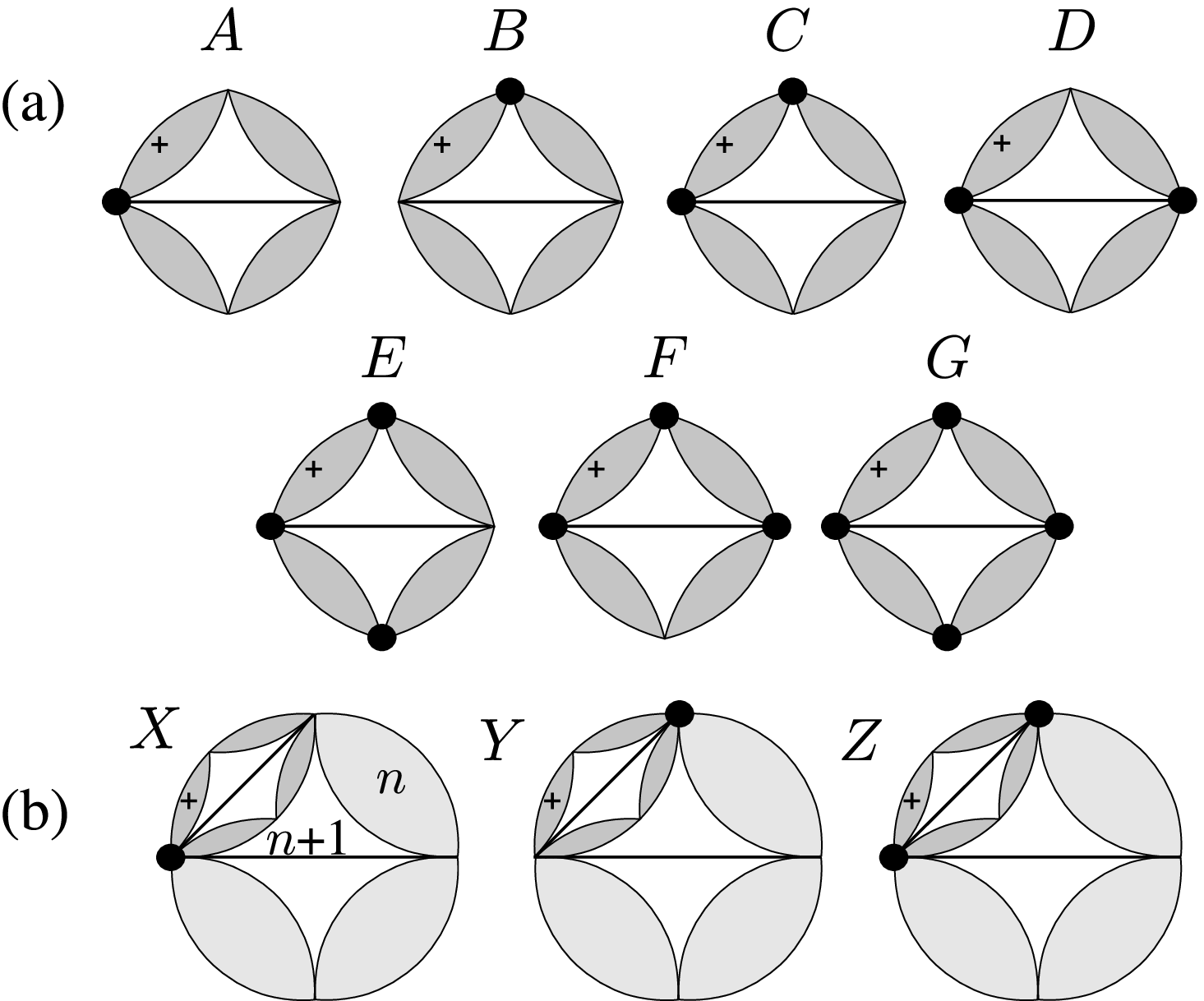}
\caption{Contact probabiliies used in the derivation of Eq.~(\ref{AtoG}).
}
\label{fig10}
\end{figure}

The recursion relations for the giant component are slightly more involved than in previous cases.
We define, as usual, $A,B,\dots,G$ as the probabilities that a node reaches various hubs combinations
in generation $n$ (Fig.~\ref{fig10}a).  We also denote by $X$ the probability that, after embedding the $n$-th generation in generation $n+1$, the node reaches only one of the hubs, connected to the non-iterated link. Similarly,
$Y$ is the probability that it reaches a single hub that is {\it not\/} connected to the non-iterated link, and
$Z$ the probability that it reaches both hubs (Fig.~\ref{fig10}b).
We then have
\begin{equation}
\label{AtoG}
\begin{split}
&A'=XqQ\,,\\
&B'=YQ\,,\\
&C'=(X+Y)qPQ+ZqQ^2\,,\\
&D'=XpQ^3\,,\\
&E'=YqP^2Q+ZqPQ^2\,,\\
&F'=X\{pPQ^2+Q[qP^2+p(2PQ+P^2)]\}\\
&\>\>\>\>\>\>\>\>\>+YpPQ^2+ZQ^2(qP+p)\,,\\
&G'=(X+Y)P[p(2PQ+P^2)+qP^2]\\
&\>\>\>\>\>\>\>\>\>+Z(1-qQ^2)\,,
\end{split}
\end{equation}
where
\begin{equation}
\begin{split}
&X=\frac{1}{2}(A+C+E)\,,\\
&Y=\frac{1}{2}(A+C+E)\,,\\
&Z=D+F+G\,.
\end{split}
\end{equation}
Again, the fact that $A'+B'+\cdots+G'=X+Y+Z$ confirms that the equations are consistent. 

The recursion equations simplify with the substitution $S\equiv X=Y$ and $T\equiv Z$, leading to
\begin{equation}
\left(
\begin{array}{c}
S'\\
T'
\end{array}\right)=\left(
\begin{array}{cc}
\frac{1}{2}qQ(1+P)^2 & \frac{1}{2}qQ^2(1+P)\\
P^2+pQ(1+P)&1-qQ^2(1+P)
\end{array}\right)
\left(
\begin{array}{c}
S\\
T
\end{array}\right).
\end{equation}
The scaling of $\Pinf$ near the percolation threshold is dominated by the largest eigenvalue of the 
recursion matrix, $\lambda=(3+\sqrt{6})/6$ (evaluated at $p_c$).  From this we find the finite size exponent
$\theta=-\ln\lambda/\ln4=0.0694207\dots$

In order to compute the order parameter critical exponent, $\beta$, we must first replace the scaling relations suitable for fractals and regular spaces with relations for percolation in {\it small world\/}
substrata, that are {\it infinite\/}-dimensional.  Instead of~(\ref{Pscaling_frac}) we write
\begin{equation}
\label{newPhi}
P=\Phi\Big(\Delta p\,N^x\Big),
\end{equation}
where the original scaling argument $L^{1/\nu}$ has been changed to  $N^x$, obviating the question of diameter.  Using this and a similar argument to the one leading to~(\ref{thetanud}), we now
derive
\begin{equation}
x=\frac{\theta}{\beta}\,.
\end{equation}
Naively equating this relation to~(\ref{thetanud}) one obtains $x=1/\nu d$.  This is, of course, meaningless, but makes some kind of sense: because the net is small world both its
dimension $d$ and  the inverse of the correlation length exponent $1/\nu$ are infinite, but in such a way that their ratio yields a finite $x$.

\begin{figure}[ht]
  \includegraphics*[width=0.40\textwidth]{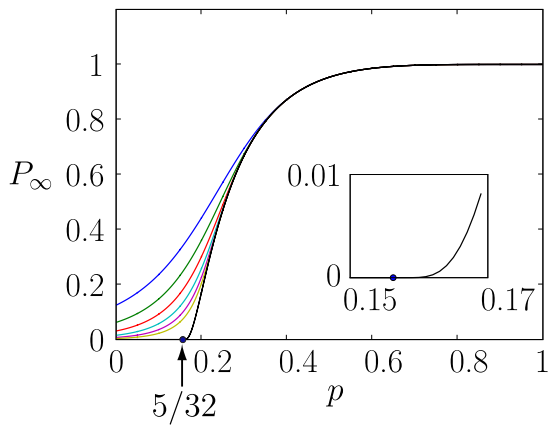}
\caption{(Color online) $\Pinf$ for the decorated $(2,2)$-flower.  Shown are curves for generations $n=6,7,8,9,10,11,$ and
$\infty$.  Inset: Detail about $p=p_c$, showing that $\Pinf\sim(p-p_c)^{\beta}$ with $\beta=\infty$.
}
\label{fig11}
\end{figure}

The exponent $x$ is obtained in practice from the recursion relation for the contact probability, 
and using~(\ref{newPhi}):
$(N'/N)^x=\Lambda=\partial P'/\partial P|_{p=p_c}$.  For the decorated $(2,2)$-flower we find $x=0$,
so that $\beta=\infty$.  Iterated curves of  $\Pinf$ (for $n\to\infty$) indeed show a transition at $p_c=5/32$ with an infinite order parameter exponent $\beta$ (Fig.~\ref{fig11}).   

\section{DISCUSSION}
\label{discussion}

We have studied the percolation phase transition in a class of hierarchical scale-free nets that can
be built to display a large variety of structural properties and that can be analyzed exactly.  When the scale-free nets are also fractal, that is, when the mass of the net increases as a power of its diameter,
percolation is very similar to what is found in regular lattices.  We do not see any specific signature
that might be ascribed to the scale-free degree distribution.

Percolation in {\it small world\/} hierarchical lattices is more exotic.  In the $(1,v)$-flowers, we find that there is no percolation phase transition: the system is always in the percolating phase, even as the bonds get diluted to concentration $p\to0$.  This is in line with what is known for stochastic scale-free
nets of degree exponent $\gamma<3$.  However, for $(1,v)$-flowers the percolation phase transition
fails to appear even as $v$ (and $\gamma$) increase without bound.  To be sure, $\Pinf(p)$ flattens
more pronouncedly about the origin, reaching near zero probability at wider and wider regions of $p$,
as $v$ increases, but there is no transition nevertheless --- see Eq.~(\ref{essential}).  A possible
cause for the exceptional resilience of $(1,v)$-flowers is their being strongly assortative, compared to
stochastic and everyday life scale-free nets.

We then studied percolation in the decorated $(2,2)$-flower, a hierarchical scale-free net of degree 
exponent $\gamma=3$ that is small world and {\it disassortative\/}.  In this case there is a percolation
phase transition at a finite $p_c$, and the order parameter critical exponent characterizing the transition
is $\beta=\infty$.  This agrees with the result for percolation in stochastic scale-free nets, that $\beta=
1/(\gamma-3)$~\cite{cohen2}, since the degre exponent of the decorated $(2,2)$-flower is $\gamma=3$. 
However, the finding in our case is generic:  one can
show that $\beta=\infty$ for decorated flowers with other $1<u\leq v$ values,  
independently of $\gamma$.  The same is true for the $(2,2)$-flower decorated with {\it two\/} non-iterated links (connecting the two pairs of opposite hubs).  The decorated $(2,2)$-flower and similar 
constructs closely mimic everyday life stochastic scale-free nets (small world, disassortative, and high degree
of clustering).  Why is it then that they cannot reproduce a phase
transition with finite $\beta$?  --- Perhaps we are still missing out on some crucial structural
property, common to everyday life stochastic networks.  Another possibility is that it is a consequence
of the hierarchical flowers being {\it finitely
ramified\/} (they can be disjointed by removing a finite number of nodes, regardless of the graphs' sizes).
We do not know whether finite ramification is typical of everyday life networks.
Finding out the answers to these questions will shed further light on the structure of the complex nets around us.  

\acknowledgments
We thank James Bagrow for discussions and help with coding the $(u,v)$-flowers, and Nihat Berker, Vladimir Privman, and Bob Ziff for  help with background
material and fruitful discussions.
Partial support from NSF award PHY0555312 (DbA) is gratefully acknowledged.


\begin{thebibliography}{99}
\bibitem{remark} More commonly, the scaling relation (\ref{dg}) is written as $d_f=d-\beta/\nu$,
where $d_f$ is the fractal dimension of the incipient infinite cluster and $d$ the dimension of
the substrate.
\end{thebibliography}
\end{document}